\documentclass[final,5p,times,twocolumn]{elsarticle}
\usepackage{amsbsy}
\usepackage{amssymb}
\usepackage{amsmath}
\newcommand{\beq}{\begin{equation}}
\newcommand{\eeq}{\end{equation}}
\newcommand{\be}{\begin{eqnarray}}
\newcommand{\ee}{\end{eqnarray}}

\journal{Astroparticle Physics}

\begin{document}
\begin{frontmatter}
\title{Propagation of high-energy neutrinos in a background of ultralight scalar dark matter}
\author[label1]{Mat\'{\i}as M. Reynoso}
\author[label1]{Oscar A. Sampayo}

\address[label1]{IFIMAR (CONICET-UNMdP) and Departamento de F\'{\i}sica, Facultad de Ciencias Exactas y Naturales, Universidad Nacional de Mar del Plata, Funes 3350, (7600) Mar del Plata, Argentina}
\begin{abstract}
{If high-energy neutrinos propagate in a background of ultralight scalar field particles of dark matter ($m_\varphi \sim 10^{-23}$eV), neutrino-dark matter interactions can play a role and affect the neutrino flux. In this work we analyse this effect using transport equations that account for the neutrino regeneration as well as absorption, and we consider the neutrino flux propagation in the extragalactic medium and also through the galactic halo of dark matter. We show the results for the final flux to arrive on Earth for different cases of point and diffuse neutrino fluxes. {We conclude that this type of neutrino interactions with ultralight scalar particles as dark matter can yield very different results in the neutrino flux and in the flavor ratios that can be measured in neutrino detectors such as IceCube.}}
\end{abstract}
\end{frontmatter}

\section{Introduction}\label{sec:intro}
The nature of dark matter (DM) remains an open issue and there are many proposals that have been made in the last years, which include: the well spread possibility of WIMPs (e.g.\cite{wimps}), superheavy particles \cite{superheavy1}, primordial black holes \cite{primordialbh}, sterile neutrinos \cite{sterilenu}, axions \cite{axions}, and also other scalar field particles \cite{scalardm}. In particular,  ultralight scalar particles with a mass $m_\varphi\sim 10^{-23}$eV \cite{uldm1,uldm2} have been argued to be a viable candidate as dark matter constituent, avoiding the overproduction of both substructure in the galactic haloes and  satellite dwarf galaxies that are not observed but are normally predicted within standard cold DM models \cite{uldm1,subexcess,dwarfexcess}. 

In the present work, we focus on what could be the consequences for a high-energy flux of neutrinos ($\nu_\alpha,  \alpha=\{e,\mu,\tau\}$) if they propagate in a universe in which DM is primarily composed of such ultralight scalar field particles ($\varphi$). In a previous work \cite{barranco}, the attenuation of a neutrino flux via interactions with this type of particles was considered neglecting the regeneration effect and the flipping of mass eigenstates which are actually unavoidable if such interactions occurred. Here perform a detailed calculation taking into account the mentioned effects by solving a system of transport equations that describe the evolution of neutrinos as they propagate in the extragalactic medium, considering also the effects of expansion of the universe. We also study the neutrino propagation through the galactic halo of dark matter, which can introduce a neutrino deficit towards the direction of the galactic center and direction dependent flavor ratios. In view of the recent data by IceCube on the flavor composition \cite{icecube1,icecube2}, although more statistics is necessary, there is still significant room for departures from a standard composition $(f_e:f_\mu:f_\tau)\simeq (1:1:1)$ of the neutrinos arriving on Earth. Hence, we show here that if the neutrino mass hierarchy follows a normal ordering or an inverted one, the neutrino flavor ratios will be affected differently by the neutrino interaction with the ultralight scalar particles.

This work is organized as follows. In section 2 we compute the relevant cross sections and the optical depths for neutrino propagation. In section 3, we calculate the neutrino flux to be arrive on Earth from point neutrino sources located at different redshifts, and in section 4 we study the case of a diffuse neutrino flux. Finally in section 5, we conclude with a brief discussion.

\section{Interactions of neutrinos and ultralight scalar DM particles}

\begin{figure}[t]
\centering 
\includegraphics[width=.6\textwidth,trim=50 550 0 0,clip]{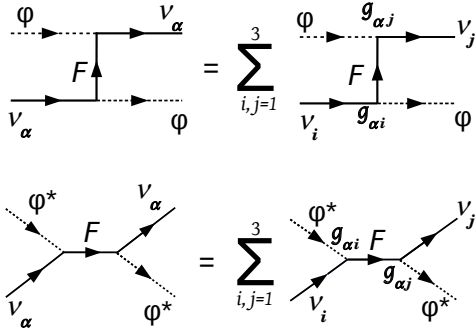}
\caption{\label{fig:diagrams} Diagrams for the $\nu_i \varphi \rightarrow \nu_j \varphi$ interactions.}
\end{figure}

The neutrino-DM interactions are in principle introduced by Lagrangian contributions for each neutrino flavor $\alpha =\{e,\mu,\tau\}$,
\be
\mathcal{L}_{\nu_\alpha\varphi}&=& g_{\alpha}\bar{\nu}_\alpha \varphi P_R F + \rm{h.c.} \label{Lflavor}, 
\ee
where $F$ represents a new fermion field with mass $M_F$ and $g_\alpha$ is the $\nu_\alpha$-$\varphi$ coupling. 
Since a neutrino of flavor $\alpha$ is a superposition of the mass eigenstates $\nu_i; \ i=1,2,3$, it is possible to rewrite the expression above as
\be
\mathcal{L}_{\nu_\alpha\varphi}= \sum_i g_i \bar{\nu}_i  \varphi P_R F + \rm{h.c.} \label{Lmass}, 
\ee
where we have introduced the couplings $ g_{i}= \sum_\alpha g_{\alpha\, i}$, with $g_{\alpha\, i}= U_{\alpha i} g_{\alpha}$ for each neutrino mass eigenstate with the scalar \cite{farzanpalo}. This will be useful in the present context in order to describe the neutrino propagation, as is discussed in the next section. For the elements of the PMNS neutrino mixing matrix $U_{\alpha i}$, we assume that the mixing angles are given by \cite{pdg2014}: $\sin^2{\theta_{12}}=0.308$, $\sin^2{\theta_{13}}=0.0234$, $\sin^2{\theta_{23}}=0.437$, and $\delta= 250^\circ$ in the case of a normal hierarchy of neutrino masses (NH, $m_1<m_2<m_3$). If the mass ordering is inverted (IH, $m_3<m_1<m_2$), then $\sin^2{\theta_{13}}=0.024$ and $\sin^2{\theta_{23}}=0.455$.
We shall consider, as in Ref. \cite{barranco}, the cases of self-conjugate ($\varphi^* =\varphi$) and non-self-conjugate dark matter ($\varphi^* \neq \varphi$) and the diagrams for the possible processes are shown in fig. \ref{fig:diagrams}, where the regeneration cases are the ones with $j=i$, while in those with $j\neq i$ the neutrino mass flips form from $m_i$ to $m_j$.

As for the bounds, they have been placed on the $\nu_\alpha$-$\varphi$ couplings, $g_\alpha$, as noted in Ref. \cite{farzanpalo}. The most stringent bounds come from the decay of $\pi$ and $K$ mesons, $|g_e|^2< 10^{-5}$ and $|g_\mu|^2< 10^{-4}$ if the mass of the fermion is $M_F \ll m_{\pi,K}$, while for higher fermion masses these bounds could be avoided. Still, since the least constrained coupling is $|g_\tau|<1$, in this work we shall adopt $g_e=g_\mu=0$, and only $g_\tau\neq 0$ in order to illustrate the possible effects on the fluxes and flavor ratios of neutrinos caused by $\nu_i$-$\varphi$ interactions {through the corresponding couplings $g_i$.}

\begin{figure*}[ht]
\centering 
\includegraphics[width=0.8\textwidth,trim=0 1700 100 0,clip]{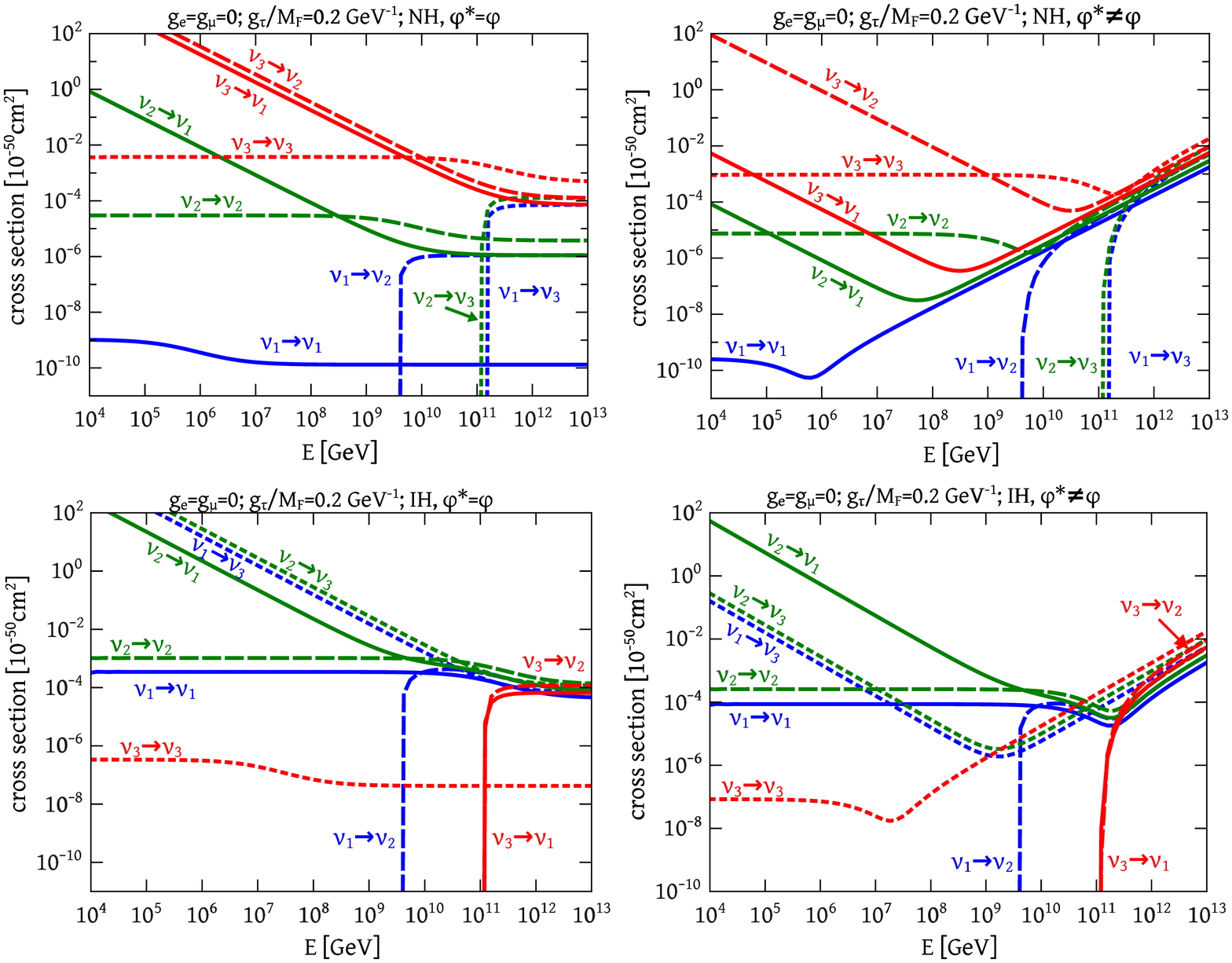}
\caption{\label{fig.sigmas} Cross sections for the processes $\nu_i \varphi \rightarrow \nu_j \varphi$. The left (right) panels correspond to (non-)self-conjugate scalar DM, and the upper (lower) panels correspond to the NH (IH) case. Blue, green, and red lines correspond to an initial $\nu_1$, $\nu_2$, and $\nu_3$, respectively. Solid, dashed, and dotted lines correspond to a final  $\nu_1$, $\nu_2$, and $\nu_3$, respectively.} 
\end{figure*}

In the case of non-self-conjugate dark matter, only the channel $u$ contributes and we obtain the following averaged squared scattering amplitude (assuming $c=1, \ \hbar=1$),
\begin{multline} 
\overline{\left|\mathcal{M}^{ij}_{\varphi*\neq\varphi}\right|^2}=
\frac{g_i^2 g_j^2}{M_F^4}\left[
  (m_3^2 - m_\varphi^2)  (p_1\cdot p_3)\right. \\ \left. + 2\,(p_1\cdot p_2) ((p_2\cdot p_3)-m_3^2)\right]
,
\end{multline}
where the $p_1$ and $p_2$ are the initial neutrino and scalar momenta, respectively, and $p_3$ is the momentum of the outgoing neutrino. We have not neglected the neutrino masses $m_i, m_j$ since they happen to play a role as we show below.  

In the case of self-conjugate scalars, the channel $s$ also contributes and we find that
\beq
\overline{\left|\mathcal{M}^{ij}_{\varphi*=\varphi}\right|^2}=
\frac{g_i^2 g_j^2}{M_{F}^4}\left[ 2\, m_i^2m_j^2+ (p_1\cdot p_3)(m_i^2+m_j^2)  \right].
\eeq

The differential cross section for the process $\nu_i \varphi \rightarrow \nu_j \varphi$ is given by
\begin{multline}
\frac{d\sigma_{ij}(E_1,E_3)}{dE}= \frac{1} {32\pi} \Theta\left[\left|\frac{E_3-\gamma E_{3,{\rm cm} }} {\gamma \beta \sqrt{E_{3,\rm cm}^2- m_i^2} }\right|-1\right] \\ \times
\frac{\overline{|\mathcal{M}^{ij}|^2}} { \sqrt{s}\gamma_{\rm cm}\beta_{\rm cm} \sqrt{E_{3,\rm cm}^2- m_i^2}
\sqrt{E_{3,\rm cm}^2- m_i^2+ m_\varphi ^2 }+ E_{3,\rm cm}} ,
\end{multline}
where:  
\be s&=& m_\varphi^2+m_i^2+ 2E_1 m_\varphi, \\
 E_{3,\rm cm}&=&(s+m_i^2-m_\varphi^2)/(2\sqrt{s}), \\
 \gamma&=&\frac{E_1 + m_\varphi}{\sqrt{s}}, \ \ \beta= \sqrt{1-\frac{1}{\gamma^2}}.
\ee  

The total cross section for the mentioned process can obtained as
\beq
\sigma_{ij}(E_\nu)= \int_{E_{3 {\rm min}}}^{E_{3 {\rm max}}} dE_3 \frac{d\sigma_{ij}(E_1,E_3)}{dE},
\eeq 
with $E_{3,{\rm min(max)}}=  \gamma\left[E_{3,{ \rm cm}}-(+)\beta \sqrt{E_{3,{\rm cm}}^2-m_i^2}\right]$. We show in fig. \ref{fig.sigmas} the obtained cross sections for the different initial and final mass eigenstates and for self-conjugate and non-self-conjugate scalar fields. We have assumed in this plot a normal hierarchy or ordering for the neutrino masses (NH) in the upper panels: $m_3=0.05{\rm eV}$, $m_2=8.7\times 10^{-3}{\rm eV}$ and $m_1= 10^{-2}m_2$ for the lightest neutrino. In the lower panels, we show the cases of an inverted hierarchy (IH), assuming $m_3=10^{-2}m_1$ with $m_1\simeq m_2= 0.049$ eV \cite{pdg2014}.

As it can be seen in fig. \ref{fig.sigmas}, the cross sections are very different for the different mass eigenstates in both the self-conjugate and non-self-conjugate cases. This is because we have kept the full scattering amplitudes without neglecting the neutrino masses, which happen to play a very important role within the kinematic regime that corresponds to the present scenario. Still, our results for the cross sections would agree with those of Ref. \cite{barranco} if the mass of the scalar was $\sim 10^{-18}$ eV and if the final and initial neutrinos have equal masses, as in their case. We note that, apart from the different masses, a different coupling corresponds to each neutrino mass eigenstates and this also explains why the cross sections are different for each of them. 

\section{Neutrino flux propagation}

Here we consider the propagation of neutrinos in the extragalactic space and also in the galactic halo.
The target of extragalactic dark matter at a given redshift $z$ is supposed to present a comoving number density given by 
\be 
 n^{\rm (eg)}_\varphi(z)= \left[\frac{3H_0^2}{8\pi G}\right] {\Omega_{\rm DM}}{\left(\frac{m_\varphi}{\rm g}\right)^{-1}} (1+z)^3,
\ee
with $H_0= 67.3 \ {\rm km \ s}^{-1}{\rm Mpc}^{-1}$, $\Omega_{\rm DM}= 0.26$, and $m_\varphi$ expressed in grams. As for the DM component of the galactic halo, the corresponding number density can be described by a NFW profile \cite{nfw} or by the proposed by Einasto et al. \cite{einasto}:
\be 
 n^{\rm (NFW)}_\varphi(r)&=& \frac{0.3 {\rm GeV \ cm}^{-3}}{ \left(\frac{m_\varphi}{\rm GeV}\right)\frac{r}{R_0}\left(1+\frac{r}{R_0}\right)^2} \\
 n^{\rm (E)}_\varphi(r)&=& \frac{7.2 \times 10^{-2}{\ \rm GeV \  cm}^{-3}}{\left(\frac{m_\varphi}{\rm GeV}\right)} e^{\left\{-\frac{2}{\alpha} \left[  \left(\frac{r}{R_0}\right)^\alpha- 1 \right] \right\}},\label{EinastoDMprofile}
\ee  
where $\alpha=0.15$, $R_0= 20 \ {\rm kpc}$ and $m_\varphi$ expressed in GeV. 

\begin{figure*}[t]
\centering 
\includegraphics[width=.88\textwidth,trim=5 8 30 0,clip]{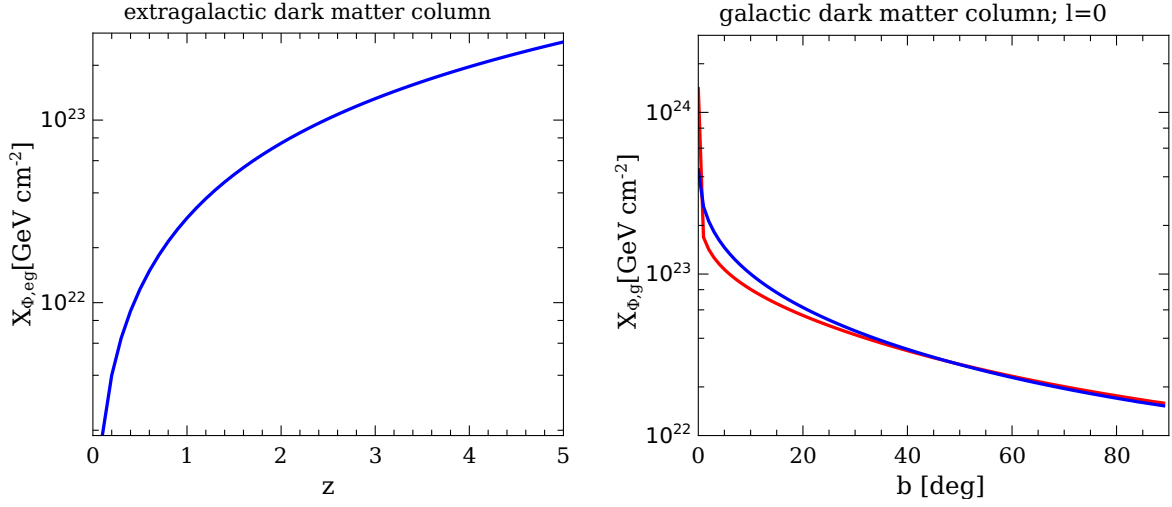}
\caption{\label{fig.xdm} Left: extragalactic dark matter column depth vs redshift. Right: Galactic dark matter column depth vs galactic latitude for $l=0$ for a NFW DM profile (red) and for the profile by Einasto et al. (blue).}
\end{figure*}

Integrating the number densities times the DM mass along the neutrino path, we can obtain the DM column densities corresponding to neutrinos propagating from a source at a given redshift $z$, $X_\varphi(z)$, and at different angles through the galactic halo $X_\varphi(l,b)$, where $l$ and $b$ are the galactic longitude and latitude, respectively. We show the results in fig. \ref{fig.xdm}. The neutrino optical depth in the case of galactic propagation can be obtained as $\tau_\nu(E)= \sigma_{\nu\varphi}(E){{X_{\varphi}(l,b)}\left(\frac{m_\varphi}{\rm GeV}\right)^{-1}}$, while for extragalactic propagation it can be obtained as 
\be  
\tau_\nu^{\rm (eg)}(E,z_{\rm s})&=& \int {dz} \, c  \left[{\frac{dz}{dt}}\right]^{-1}  n_\varphi^{\rm (eg)}(z)\sigma_{\nu\varphi}\left[E(1+z)\right], 
\ee
where $\frac{dz}{dt}= -H_0 (1+z)\sqrt{\Omega_\Lambda+\Omega_{\rm m}(1+z)^3}$, with $\Omega_{\rm m}= 0.315$.    
\begin{figure*}[tbp]
\centering 
\includegraphics[width=.95\textwidth,trim=0 52 20 0,clip]{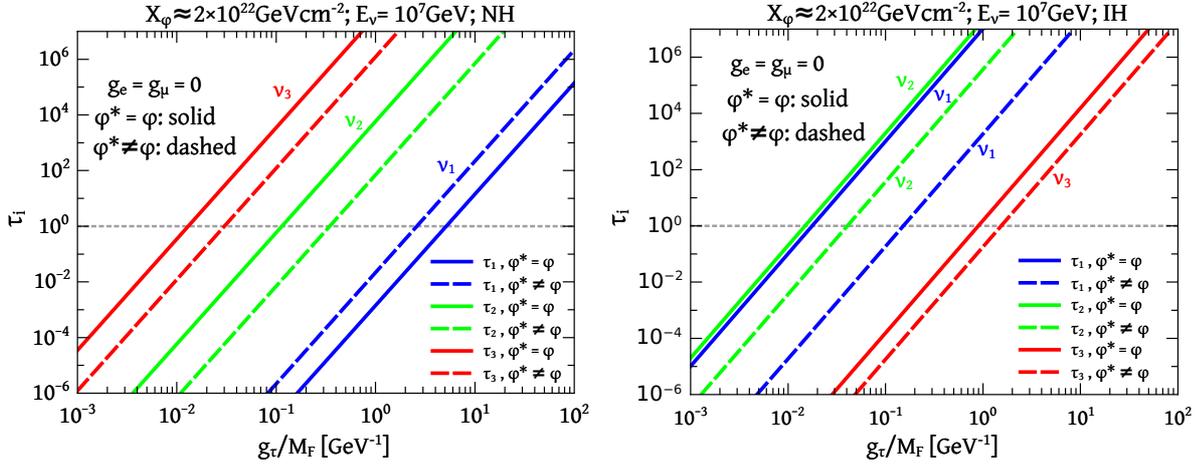}
\caption{\label{fig.opticaldepth} Optical depth for neutrinos of energy $E=10^7$ GeV if the DM column depth is $X_\varphi\approx 2\times 10^{22} {\rm GeV \ cm}^{-2}$ for non-self-conjugate (dashed lines) and self-conjugate scalar DM (solid lines), and for NH and IH in the left and right panels, respectively.}
\end{figure*}

In fig. \ref{fig.opticaldepth} we show the optical depth as a function of $\left( \frac{g_\tau}{M_F} \right)$ for a neutrino energy of $E=10^7$ GeV as an example case, for reference. In that figure, we consider a DM column depth of $X_\varphi \approx 2\times 10^{22} {\rm GeV \ cm}^{-2}$, which as can be seen from fig. \ref{fig.xdm}, corresponds to a redshift of $\sim 0.5$ or to a path going through the galactic halo at $(l,b)=(0^\circ,70^\circ)$. From fig. \ref{fig.opticaldepth} it can be seen that in order to have significant $\nu$-$\varphi$ interactions in the propagation, {it should hold that $\left( \frac{g_\tau}{M_F} \gtrsim 0.02	 {\rm GeV}^{-1}\right)$ in the self-conjugate case, and $\left( \frac{g_\tau}{M_F} \gtrsim 0.05 {\rm GeV}^{-1}\right)$ in the non-self-conjugate case}. In these cases, the heaviest neutrino faces an optical depth greater than one, and hence it would affect the flux of the different flavor neutrinos to arrive on Earth at $E=10^7$ GeV.

We now describe how we treat the neutrino propagation over cosmological distances, but easily adaptable to the case of propagation through the galactic halo. 
The processes $\nu_i\varphi \rightarrow \nu_j\varphi$ include regeneration ($j=i$) and flipping processes ($j\neq i$). The latter processes represent absorption effects for the flux of $\nu_i$ neutrinos and also implies a re-injection effect of $\nu_j$ neutrinos.
In terms of the comoving neutrino density for each mass eigenstate $N_{\nu_i}$, we can write 
\begin{multline} 
 \frac{\partial N_{\nu_i}}{\partial t}=Q_{\nu_i} - 3H\, N_{\nu_i} +\frac{\partial\left[ H \ E \ N_{\nu_i} \right]}{\partial E} 
-N_{\nu_i} \sigma_{{\nu_i}\varphi,{\rm tot}}(E) c n_\varphi \\ + \sum_{j=1,2,3}\int_E^\infty dE'\frac{d\sigma_{ji}(E',E)}{dE} c \ n_\varphi N_{\nu_j}(E',t),
\end{multline}
which on the right member includes the neutrino injection by the source $Q_{\nu_i}$, the effects of expansion of the universe in the second and third terms, absorption in the fourth term, and regeneration and re-injection in the last term \cite{gazivenia}. Here, the total cross section is denoted by $\sigma_{{\nu_i}\varphi, {\rm tot}}= \sum_{j=1,2,3} \sigma_{ij}$.
In the present case, the differential cross sections $\frac{d\sigma_{ij}(E',E)}{dE}$ are very sharp functions of $E'$ which are significant only for $E'\simeq E$. This means that the energy loss in each interaction is very small. Hence, we can treat the process as a continuous loss one, in which case we have
\begin{multline} 
 \frac{\partial N_{\nu_i}}{\partial t}=Q_{\nu_i} - 3H\, N_{\nu_i}+\frac{\partial\left[ H \ E \ N_{\nu_i} \right]}{\partial E}
-N_{\nu_i} \sigma_{{\nu_i}\varphi,{\rm tot}}(E) c n_\varphi  \nonumber \\
 + \sum_{j=1,2,3} \left\{c n_\varphi\sigma_{ji }N_{\nu_j}+\frac{\partial\left[b_{ji}N_{\nu_j}\right]}{\partial E}\right\}
\end{multline}
where
\be  
b_{ji}(E)= n_{\varphi}c\int_E^\infty dE' \frac{E^2}{E'}  \frac{d\sigma_{ji}}{dE},
\ee 
and an effective injection:
\be
Q_{\nu_i}^{\rm eff}= Q_{\nu_i}+ \sum_{j\neq i}\left\{n_\varphi c \sigma_{ji }N_{\nu_j}+\frac{\partial\left[b_{ji}N_{\nu_j}\right]}{\partial E}\right\}.\label{Qeff}
\ee

Hence, it is possible to express the transport equation in terms of redshift as
\begin{multline} 
 \frac{\partial N_{\nu_i}}{\partial z}=-\frac{Q_{\nu_i}^{\rm eff}}{H(z)(1+z)} + {N_{\nu_i}}B \\ 
 -\left(\frac{\partial N_{\nu_i}}{\partial E} \right)\left[\frac{E}{(1+z)}+\frac{b_{ii}}{H(z){(1+z)}}\right] 
\label{transport_eq_z},
\end{multline}
and the solution can be found using the method of characteristics as
\be
N_{\nu_i}(z,E)=  \int_z^{z_{\rm max}}dz'\frac{Q_{\nu_i}^{\rm eff}}{H(z')(1+z')}\exp\left[-\int_z^{z'}B(z'',E'')\right],\label{sol_transport_eq} 
\ee
where
$$B(z,E)=\left[\frac{2}{(1+z)}+ \frac{\sigma_{i\neq j }n_\varphi c }{H(z){(1+z)}}-\frac{1}{H(z){(1+z)}}\frac{\partial b_{ii}}{\partial E}\right]. $$
The flux of neutrinos of a flavor $\alpha$ at $z=0$ can be obtained as 
\be 
  J_{\nu_\alpha}(E)= \frac{c}{4\pi}\sum_{i}\left|U_{\alpha i}\right|^2 N_i(z=0,E),
\ee 

which is actually an incoherent superposition of the three mass eigenstates. Although neutrinos are emitted and detected as weak flavor eigenstates, the new interaction with the scalar selects one mass eigenstate and destroys coherence, in a similar fashion as it happens if a quantum graviton interacts with neutrinos Ref \citep{miller2013}: if the interaction takes place, then the neutrino exists in a single mass eigenstate. Hence, in the present case, the decoherence assumption is justified, in contrast to other cases in which coherence has to be kept \citep{pas2005,miranda2015,aiekens2014}. We can, then, proceed to work out the neutrino propagation following the mass eigenstates and adding them up incoherently to obtain the flavor neutrinos to be observed on Earth, as it is also done in Ref. \citep{farzanpalo}. We also consider here the propagation through the galactic DM halo of extragalactic neutrinos that have already decohered due to the interactions with the scalars even before arriving to the halo, as we show in the next section.

We have a system of three coupled equations, one eq. (\ref{transport_eq_z}) for each massive neutrino $\nu_i$. However, it is possible to solve for each $N_{\nu_i}$ at a time, taking into account the relative dominances of all the possible $\nu_i \varphi \rightarrow \nu_j \varphi$ processes. As is shown in fig.\ref{fig.sigmas}, for neutrinos with energies $E \lesssim 5\times 10^9{\rm GeV}$, in the NH case we have that neither $\nu_1$ or $\nu_2$ neutrinos produce $\nu_3$ neutrinos, which implies that the first to solve is $N_{\nu_3}$. Then, since $\nu_3$ neutrinos produce $\nu_1$ and $\nu_2$ neutrinos, and in turn, $\nu_1$ neutrinos just generate $\nu_1$ neutrinos of lower energy, the second distribution to solve is $N_{\nu_2}$. Finally, we solve for $N_{\nu_1}$ taking into account the re-injection generated by the processes $\nu_3\varphi\rightarrow \nu_1\varphi$ and $\nu_2\varphi\rightarrow \nu_1\varphi$. This procedure holds for both self conjugate and non-self-conjugate cases of DM scalar. In the IH case, a similar analysis implies that we have to solve first for $N_{\nu_2}$, second for $N_{\nu_1}$, and last for $N_{\nu_3}$, as long as we keep within the mentioned energy range, as we shall do in the present work.

\section{Application to point sources}
\begin{figure*}[t]
\centering 
\includegraphics[width=.9\textwidth,trim=0 0 0 0,clip]{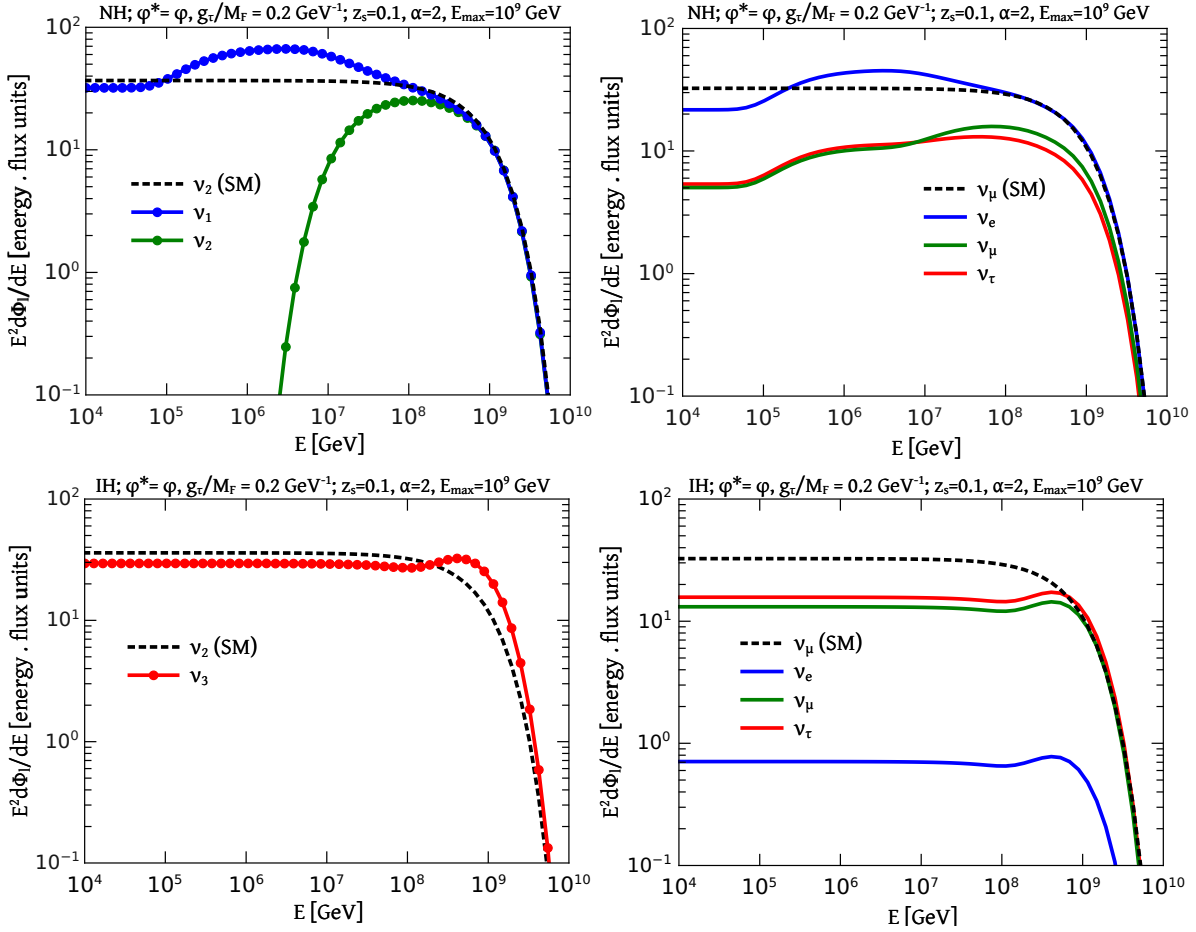}
\caption{\label{fig.flux_z01} Neutrino fluxes on our galaxy from a source at $z_{\rm s}=0.1$, with $g_{\tau}/M_F=0.2 \ {\rm GeV}^{-1}$ for non-self-conjugate and self-conjugate DM in the upper and lower panels, respectively.  The flux of neutrino mass eigenstates (left) and the flux of the different flavor of neutrinos (right) are shown. A normal ordering for the masses is assumed.}
\end{figure*} 
Here, we show some results for neutrino point sources at different redshifts $z_{\rm s}$ assuming that the injected neutrino spectrum of each neutrino flavor presents the following dependence with energy at the source:
\be
Q_{\nu}(E,z)\propto \delta \left[z-z_{\rm s}\right] E^{-\alpha} \exp{\left(-\frac{E}{E_{\rm max}} \right)},\label{powerlawflux}   
\ee
where $E_{\rm max}$ is a break or maximum energy of the neutrinos, related to the proton acceleration mechanism operating at the particular system.

We further suppose that the flavor ratios ($f_\alpha=\frac{Q_{\nu_\alpha}}{Q_{\nu_e}+Q_{\nu_\mu}+Q_{\nu_\tau}}$) as emitted from the source correspond to $(f_e,f_\mu,f_\tau)= (\frac{1}{3}:\frac{2}{3}:0)$, as it is typically expected for neutrinos generated after pion production \footnote{Note that magnetic effects at the source \cite{lipari} or the effect of pion and muon acceleration \cite{reynoso2014,winter2014} could yield a different flavor composition at the source.}. This implies that the composition of emitted neutrino mass eigenstates is: $(f_1:f_2:f_3)\propto(|U_{e1}|^2+2|U_{\mu 1}|^2: |U_{e2}|^2+2|U_{\mu 2}|^2:|U_{e3}|^2+2|U_{\mu 3}|^2)= (0.989:1.134:0.877)$ in the NH case, and $(f_1:f_2:f_3)\propto(0.978:1.110:0.912)$ in the IH case. Thus, the injection of the corresponding neutrino mass eigenstates that we consider in eq. (\ref{Qeff}) is $Q_{\nu_i}(E,z)=f_i Q_\nu(E,z)$.

In fig. \ref{fig.flux_z01}, we show the neutrino fluxes to arrive on out galaxy from a point source at redshift $z_{\rm s}=0.1$ with $\alpha=2$ and $E_{\rm max}= 10^{9}$GeV, for neutrino-DM couplings given by $g_e=g_\mu=0$ and $g_\tau/M_F= 0.2 \  {\rm GeV}^{-1}$. In this case, we consider the case of self-conjugate DM, but qualitatively similar results are obtained in the non-self-conjugate case {for lower couplings as we shall see below}. In the upper panels, we show the  fluxes in the NH case, for which the $\nu_3$ mass eigenstate is the heaviest one and hence the most rapidly affected by the interaction, as can be seen in the upper left panel where only the fluxes of $\nu_1$ and $\nu_2$ are significant. In the upper right panel, we show the corresponding fluxes of neutrinos of the three flavors, where it can be seen that the $\nu_e$ flux dominates over the $\nu_\mu$ and $\nu_\tau$ fluxes. In the lower panels, the fluxes shown correspond to the IH case, where in contrast to the NH case, it can be seen that the most affected mass eigenstates are $\nu_1$ and also $\nu_2$, and this implies the fluxes of flavor neutrinos dominated by the $\nu_\tau$ and $\nu_\mu$ as seen in the lower right panel.

\begin{figure*}[t]
\centering 
\includegraphics[width=0.95\textwidth,trim=30 0 0 0,clip]{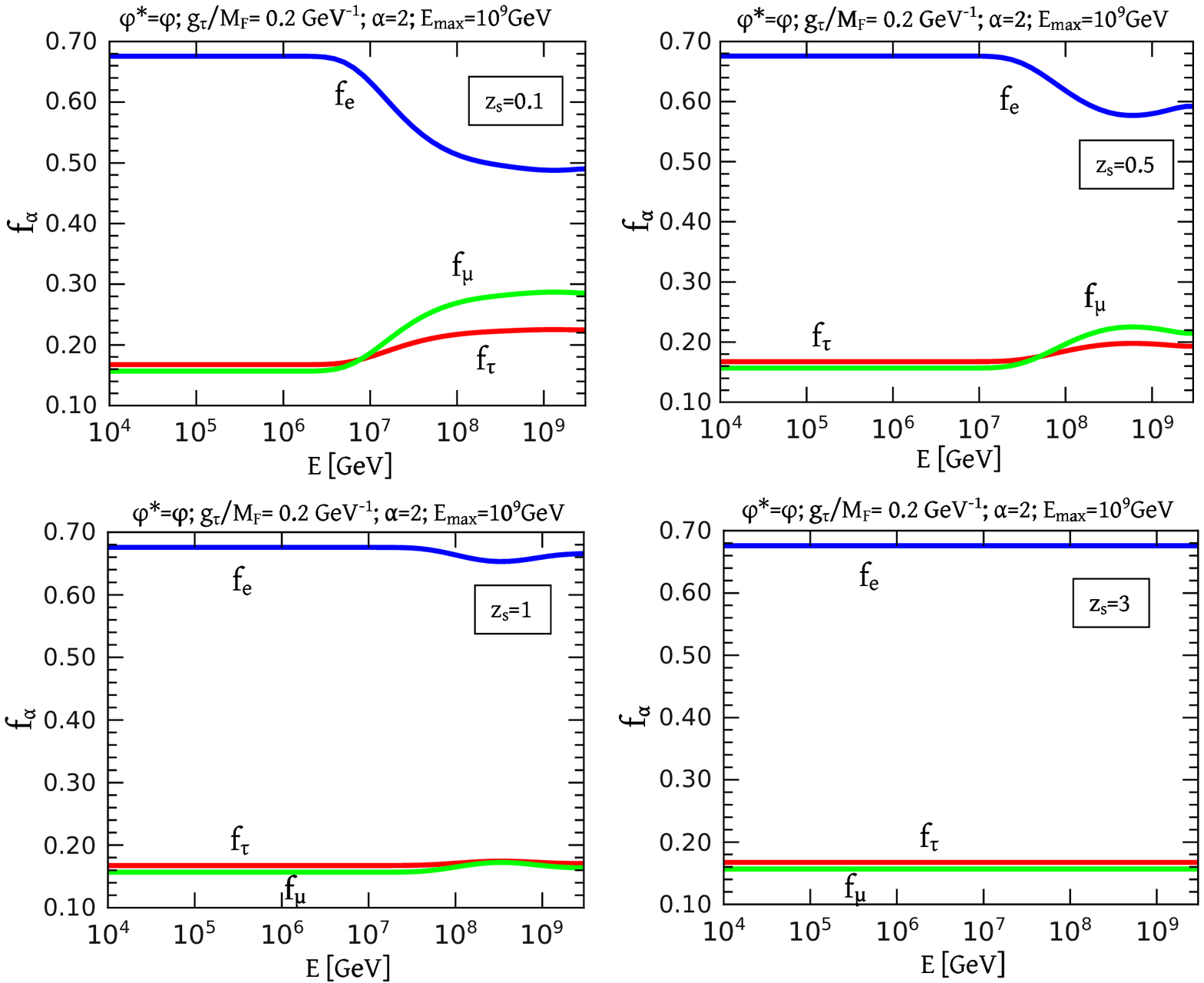}
\label{fig.ratios_vs_E} \caption{Neutrino flavor ratios on our galaxy from sources at resdshfts $z_{\rm s}=0.1, 0.5, 1,$ and $3$, with $\alpha=2$ and $E_{\rm max}=10^9$ GeV. Here we have assumed $g_e=g_\mu=0$, $g_{\tau}/M_F=0.2 \ {\rm GeV}^{-1}$, self-conjugate DM, and NH for the neutrino masses.}
\end{figure*} 

 We show in fig. \ref{fig.ratios_vs_E}, for the self-conjugate case, the neutrino flavor ratios $f_{\alpha}$ after extragalactic propagation from point sources at different redshifts, i.e., before going through the galactic halo. We observe here that as more DM is traversed, the flavor ratios vary. Since in the NH case the $\nu_1$ flux is the least affected one of the three fluxes of mass eigenstates, in the limit case only this flux would survive, which would lead to a flavor ratio composition $(f_e:f_\mu:f_\tau)= (|U_{e1}|^2:|U_{\mu 1}|^2:|U_{\tau 1}|^2)\simeq (0.67:0.16:0.17)$. This can be seen for instance in the bottom right panel of fig. \ref{fig.ratios_vs_E} for $z_{\rm s}=3$ where the coupling with DM is quite strong, $g_\tau/M_F= 0.2 \  {\rm GeV}^{-1}$.

\begin{figure*}[tbp]
\centering 
\includegraphics[width=.83\textwidth,trim=0 50 0 0,clip]{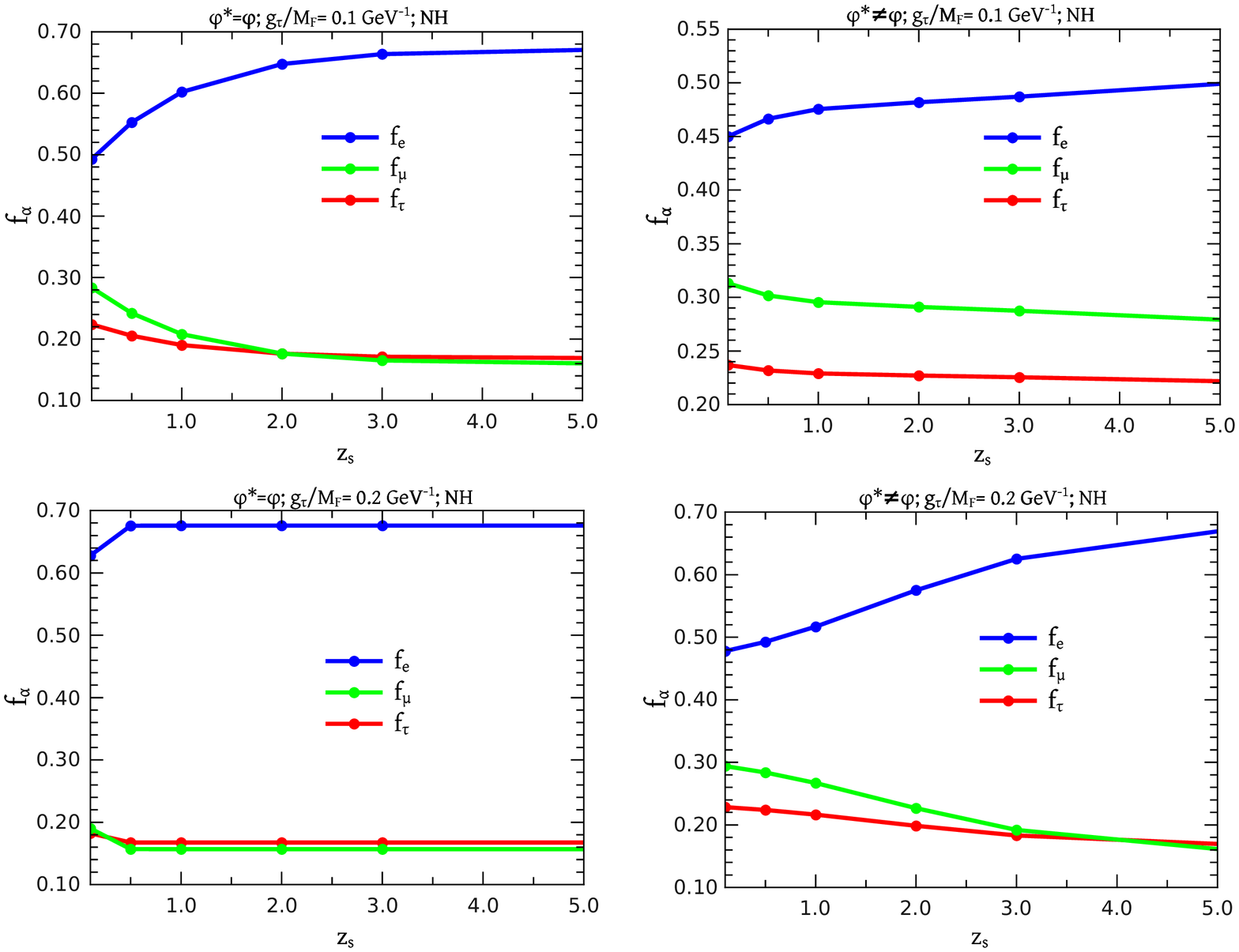}
\caption{\label{fig.ratios_vs_z_NH}Neutrino flavor ratios on our galaxy as a function of the source redshift in the NH case, for couplings $g_{\tau}/M_F=0.1 \ {\rm GeV}^{-1}$ and $g_{\tau}/M_F=0.2 \ {\rm GeV}^{-1}$ in the upper and lower panels, respectively.}
\centering 
\vspace*{5mm}
\includegraphics[width=.80\textwidth,trim=0 35 0 0,clip]{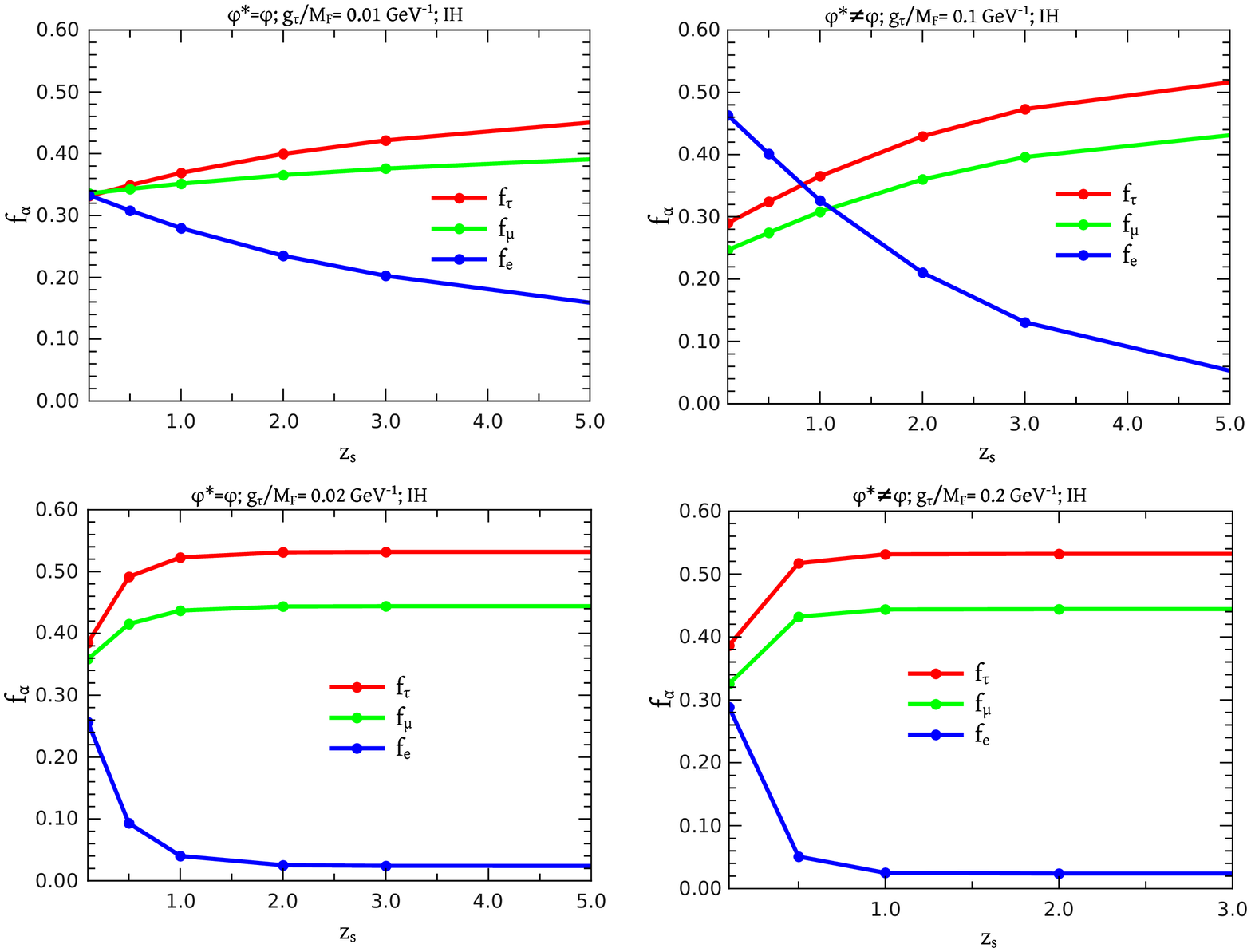}
\caption{\label{fig.ratios_vs_z_IH} Neutrino flavor ratios on our galaxy as a function of the source redshift in the IH case. {Here, $g_{\tau}/M_F=0.1 \ {\rm GeV}^{-1}$ and $g_{\tau}/M_F=0.2 \ {\rm GeV}^{-1}$ for $\varphi^*\neq \varphi$ (right-upper and right-lower panels, respectively), and $g_{\tau}/M_F=0.01 \ {\rm GeV}^{-1}$ and $g_{\tau}/M_F=0.02 \ {\rm GeV}^{-1}$ for $\varphi^*= \varphi$ (left-upper and left-lower panels, respectively).}}
\end{figure*} 

We can now take a fixed neutrino energy, $E=10^7$ GeV for instance, and plot the flavor ratio expected on our galaxy as a function of the redshift of the source, for the same emitted neutrino flux. We show such plot in fig. \ref{fig.ratios_vs_z_NH} for the NH case, and in fig. \ref{fig.ratios_vs_z_IH} for the IH case.

In these figures, we have assumed two different cases for the $\nu_\tau$-$\varphi$ coupling in order to appreciate its effect on the flavor ratios: a relatively weak coupling case in the upper panels and a stronger coupling case in the lower ones. It can be seen that the flavor ratios, as the fluxes arrive on to our galaxy, depend on both the coupling and the amount of DM traversed. Thus, for instance in fig. \ref{fig.ratios_vs_z_NH} for $\varphi^*\neq \varphi$, the flavor ratios corresponding to a very far source with $z_{\rm s}= 5$  for $g_\tau/M_F=0.1\,{\rm GeV}^{-1}$ are very similar to those corresponding to a much closer source ($z_{\rm s}\simeq 0.5$) if the coupling was twice as much. As for the IH case, we note that since in this case the lightest neutrino eigesatate and the least affected by least affected one is $\nu_3$, then the flavor composition in this case tends to $(f_e:f_\mu:f_\tau)= (|U_{e3}|^2:|U_{\mu 3}|^2:|U_{\tau 3}|^2)\simeq (0.024:0.444:0.532)$, as can be seen in fig. \ref{fig.ratios_vs_z_IH}, in particular in the lower panels for a strong coupling. In the case of a weaker coupling, we see that in the non-self-conjugate cases, the effects on extragalactic propagation is more gradual since the cross sections are lower than in the self-conjugate case for the same coupling. Also, as can be seen from figs.  \ref{fig.ratios_vs_z_NH} and fig. \ref{fig.ratios_vs_z_IH}, the overall effect on the neutrino fluxes is more significant than in the NH case. This is because in the former case, the mass of the lightest eigenstate is higher and hence more sensible to the interaction than in the NH case. {In the particular case of IH and $\varphi^*=\varphi$, it also happens that the eigenstates $\nu_1$ and $\nu_2$ are similarly affected by the interaction, which leads to a more rapid attenuation of the  fluxes.}

\begin{figure*}[t]
\centering 
\includegraphics[width=.9\textwidth,trim=0 0 0 0,clip]{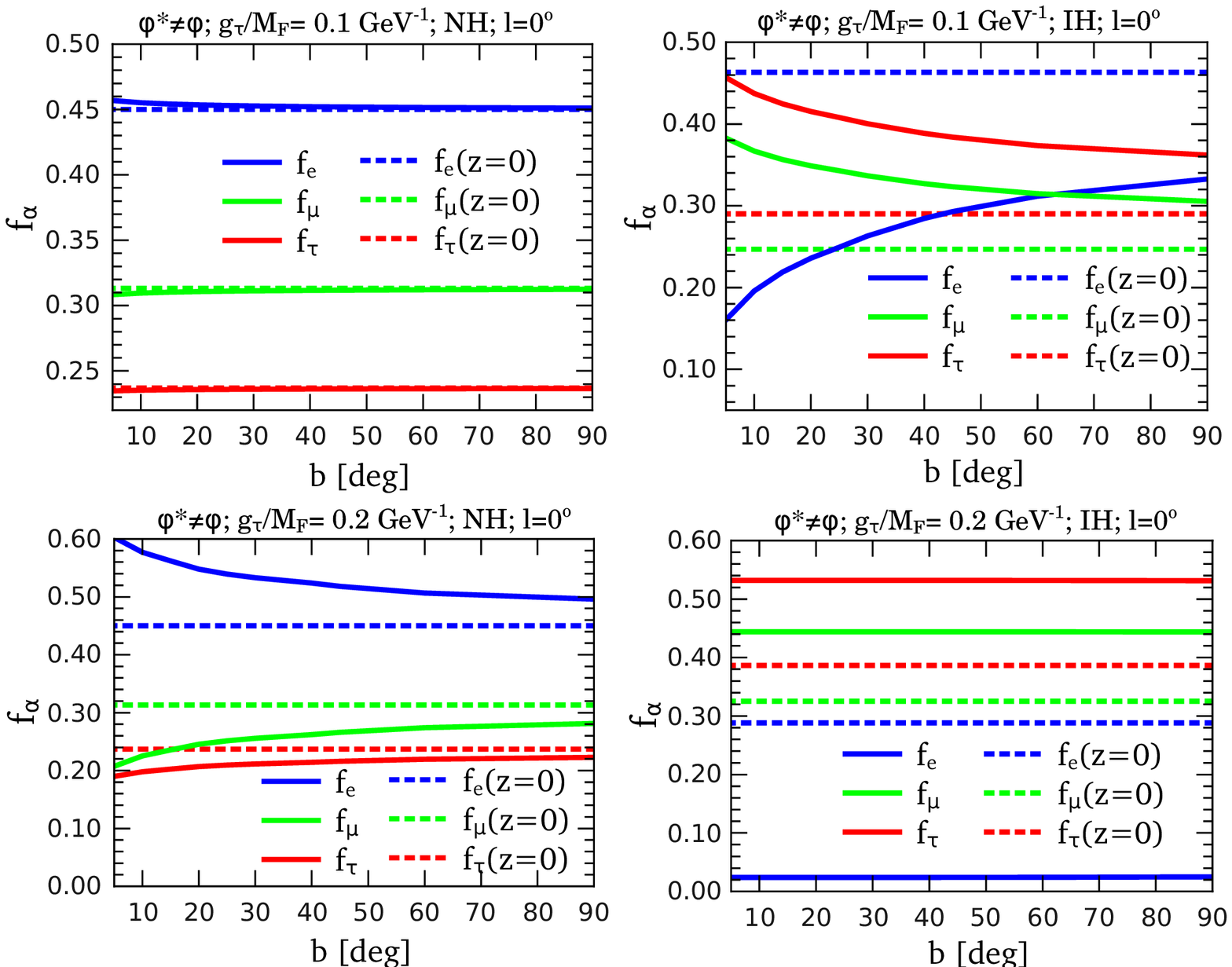} 
\caption{\label{fig.ratios_vs_b_z01_NH_IH} Neutrino flavor ratios on Earth as a function of the galactic latitude $b$ for $l=0^\circ$ and $z_s=0.1$ in the inverted hierarchy case.}
\end{figure*}

Once the neutrino fluxes arrive at the border of our galaxy, it is necessary to consider the propagation through the galactic DM halo at different possible directions on the sky corresponding to different DM column depths. We adopted in our calculation the Einasto density profile (\ref{EinastoDMprofile}), which is non-singular at the galactic center. For illustration, we show in fig. \ref{fig.ratios_vs_b_z01_NH_IH} the flavor ratios as a function of the galactic latitude $b$, at a fixed galactic longitude $l=0^\circ$ assuming fluxes emitted from a source at redshift $z_{\rm s}=0.1$. Here, again we consider the non-self conjugate case, and we show the obtained flavor ratios for the NH and IH case in the left and right panels, respectively. The upper panels correspond the weak coupling case, with $g_\tau/M_F=0.1 \, {\rm GeV}^{-1}$, and the lower panels correspond to the stronger coupling considered, $g_\tau/M_F=0.2 \, {\rm GeV}^{-1}$. In the upper left panel, there is practically no difference in all the sky in the observed flavor ratios, although they are very different from the standard expectation $(f_e:f_\mu:f_\tau)\sim (1:1:1)$. The reason for is that for the particular value of the coupling, it is only the heaviest eigenstate the one that is strongly suppressed by the interactions with extragalactic DM, while the rest of the eigenstates remain mostly unaffected even when traversing higher columns of DM through the galactic halo. The situation corresponds to optical depths for each neutrino eigenstate very similar to the ones shown in the left panel of fig. \ref{fig.opticaldepth}: only the red dashed curve, i.e. for $\nu_3$, is above 1 for a coupling $g_\tau/M_F=0.1 \, {\rm GeV}^{-1}$. If the coupling is increased by a factor of $2$, then the eigenstate $\nu_2$ begins to be affected differently in different directions through the galactic DM halo, and this leads to a variation of the flavor ratios to be observed as shown in the lower left panel of fig. \ref{fig.ratios_vs_b_z01_NH_IH}. A different situation arises in the IH case, for which even for $g_\tau/M_F=0.1 \, {\rm GeV}^{-1}$ the flavor ratios vary over different directions in the sky. This is because the masses are different than in the NH case and the optical depth for the heaviest eigenstate ($\nu_2$) is a bit lower than for $\nu_3$ in the NH case, but still much greater than one, while the second eigenstate to be affected ($\nu_1$) is also quite massive and has a higher optical depth than the one for $\nu_2$ in the NH case. This also implies than if the coupling is stronger, as in the lower right panel, then both $\nu_1$ and $\nu_2$ are almost completely suppressed in all directions in the sky accross the DM halo, and the flavor ratios to be observed on Earth correspond to the unaltered propagation of the $\nu_3$ flux, the lightest eigenstate in the IH case.%

\section{Application to a diffuse neutrino flux}

In this section, we present our results for a diffuse neutrino flux in the case that $\nu$-$\varphi$ interactions are significant. %
The possible astrophysical sources could be for instance, gamma-ray bursts, or active galactic nuclei, but we do not specify here the nature exact of the source. Rather, we consider a typical power law emitted spectrum as that of Eq. \ref{powerlawflux}, and assume that the emitting sources are distributed over redshift as the star formation rate is \cite{yuksel,wang}):
\be 
W_{\rm SFR}(z)\propto
\left\lbrace\begin{array}{ccc}
(1+z)^{3.4}  & \mbox{for} \ z<1\\ 
(1+z)^{-0.3} & \mbox{for} \ 1<z<4\\
(1+z)^{-3.5} & \mbox{for} \ z>4
\end{array}.\right. 
\ee
For illustration, in order to fix the level of neutrino flux to be observed on Earth, we use the IceCube data to normalize the standard prediction (SM), i.e., the obtained in the absence of any effects due to neutrino interactions,
\beq
J_{\nu_\alpha}^{\rm (SM)}(E) = K_{\nu_\alpha} \int_0^5dz W_{\rm SFR}(z)\left[E(1+z)\right]^{-\alpha}\exp{\left(-\frac{E(1+z)}{E_{\rm max}} \right) }.
\eeq
We adopt the power $\alpha=2.2$ and fix the constants $K_{\nu_\alpha}$ in order to satisfy for $E\ll E_{\max}=10^9{\rm GeV}$ the best fit flux by IceCube \cite{icecube2}: 
$$J_{\nu}^{\rm best\, fit}(E)=1.7\times 10^{-18}{\rm GeV^{-1}sr^{-1}s^{-1}}  \left(\frac{E}{100 {\rm TeV}}\right)^{-2.2}.$$

\begin{figure*}[tbp]
\centering 
\includegraphics[width=.9\textwidth,trim=0 0 0 0,clip]{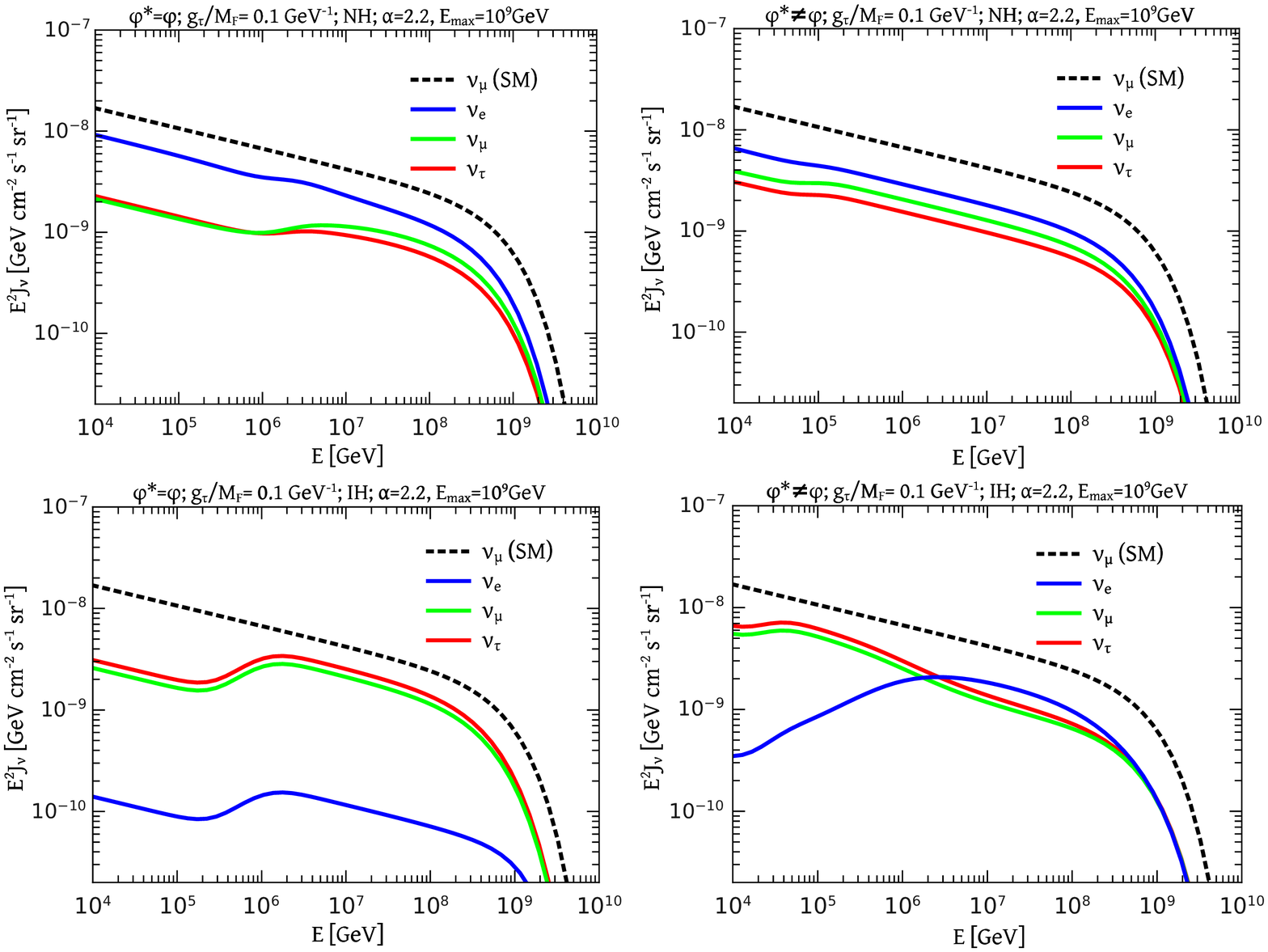}
\caption{\label{fig.fluxdiff_z0_NH_IH}Diffuse neutrino fluxes arriving on our Galaxy ($z=0$) in the NH and IH cases for $g_\tau/M_F=0.2 \ {\rm GeV}^{-1}$ in the upper and lower panels, respectively. The self-conjugate and non-self-conjugate cases are shown in the left and right panels, respectively. }
%
\centering 
\includegraphics[width=1.1\textwidth,trim=0 135 0 0,clip]{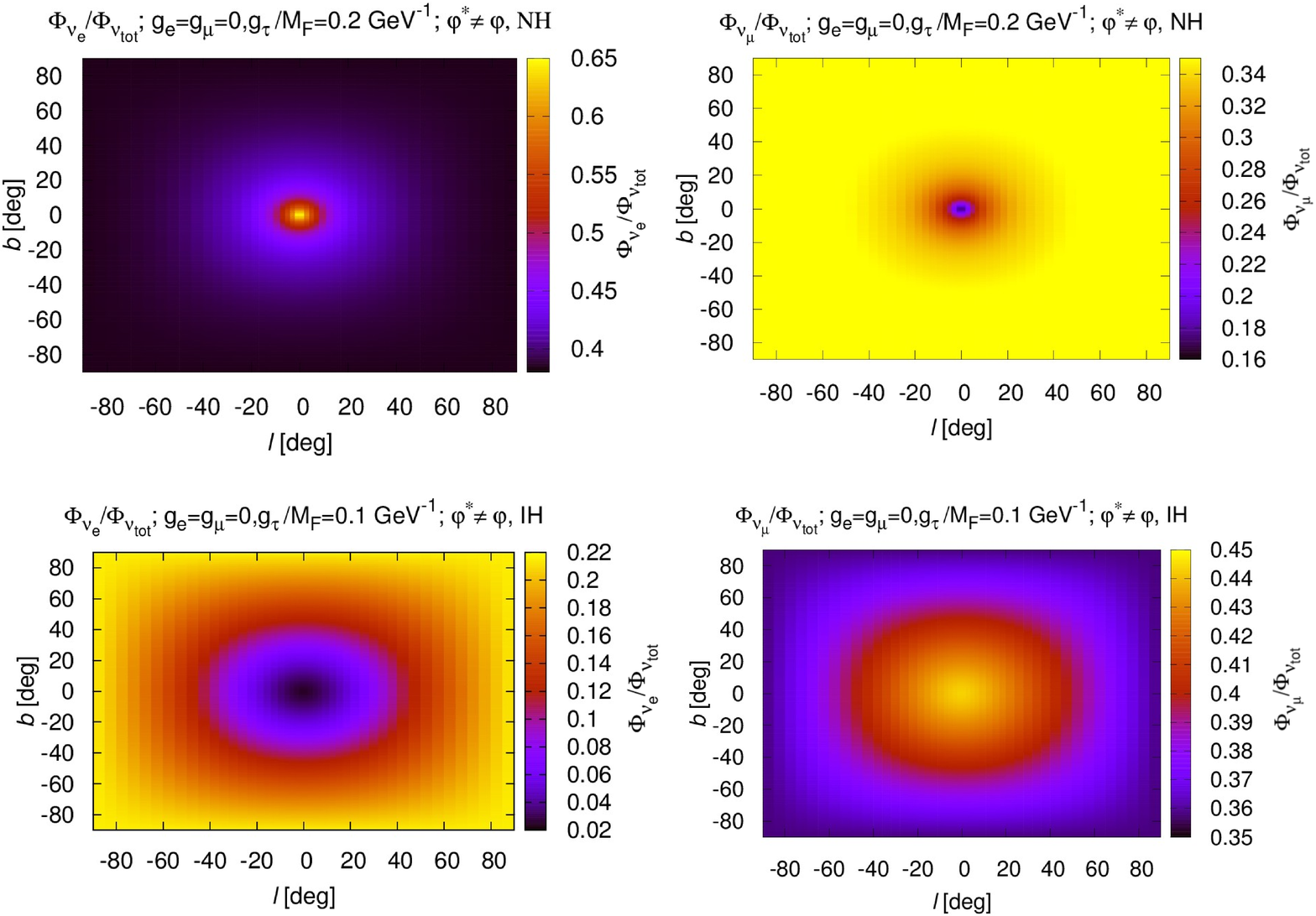}
\caption{\label{fig.fluxdiff_halo_nosc}Ratio of diffuse neutrino fluxes above $10^6$GeV, $\left(\frac{\Phi_{\nu_e}}{\Phi_{\nu_{\rm tot}}}\right)$ and $\left(\frac{\Phi_{\nu_\mu}}{\Phi_{\nu_{\rm tot}}}\right)$, arriving on Earth as a function of the galactic coordinates $(l,b)$. }
\end{figure*}

The SM $\nu_\mu$ flux is shown for reference in fig. \ref{fig.fluxdiff_z0_NH_IH} with black dashed lines, and the rest of the curves in that plot correspond to the diffuse neutrino fluxes of the different flavors that would arrive on our galaxy after being produced by a population of sources as mentioned above, and undergoing interactions with ultralight scalar DM. We show the results for the NH and the IH cases for the hierarchy of neutrino masses in the upper and lower panels, respectively. The left panels correspond to self-conjugate DM and the right panels to non-self-conjugate DM. {As can be seen, the interaction is stronger in the former case, particularly for the inverted hierarchy, as mentioned above.}  Here we note that for neutrino energies $\gtrsim 10^6{\rm GeV}$, the $\nu_e$ flux can become dominant even in the IH case, not only in the NH case. This is because, as can be seen from the lower right panel of fig. \ref{fig.sigmas}, the $\nu_1$ neutrinos that are copiously generated by the process $\nu_2\varphi\rightarrow \nu_1\varphi$ do not get absorbed so easily at those energies, and this leads to a $\nu_e$ flux higher than the corresponding to other flavors if the coupling is $g_\tau/M_F=0.1{\rm GeV^{-1}}$.
Assuming, as a first approximation, that the diffuse flux arrives isotropically on our galaxy, if $\nu$-$\varphi$ interactions are significant as in the cases of the previous plots, then the neutrino fluxes and the flavor ratios will depend on the arrival direction because of different DM column depths across the halo. To illustrate this effect, we can integrate the neutrino flux of the different flavors,
\be
\Phi_{\nu_\alpha}(l,b)=\int_{10^6{\rm GeV}}^{\infty} dE \, J_{\nu_\alpha}(E,l,b),  
\ee
and we plot in fig. \ref{fig.fluxdiff_halo_nosc} the ratios $\Phi_{\nu_\mu}/{\Phi_{\nu_{\rm tot}}}$ and $\Phi_{\nu_e}/{\Phi_{\nu_{\rm tot}}}$ as a function of the galactic coordinates $(b,l)$, in the case of non-self-conjugate dark matter and for normal and inverted hierarchy of neutrino masses (NH and IH) in the upper and lower panels, respectively. In the cases shown, the ratios $g_i/M_F$ are such that the interactions between the diffuse neutrinos and scalar DM in trajectories close to the galactic center affect primarily the two heavier mass eigenvalues, i.e. $m_2,m_3$ in the NH case and $m_1,m_2$ in the IH case. In such cases, the neutrino flux in those directions tends to be dominated by the one of the lightest mass eigenvalue, and hence the flavor composition approaches $(\frac{\Phi_{\nu_e}}{\Phi_{\rm tot}}:\frac{\Phi_{\nu_{\mu}}}{\Phi_{\rm tot}}:\frac{\Phi_{\nu_{\tau}}}{\Phi_{\rm tot}})\rightarrow (0.67:0.16:0.17)$ in the NH case, and $(0.024:0.444:0.532)$ in the IH case.

\section{Final comments}
We have studied the effects of neutrino interactions with ultralight scalar particles as the main DM constituent. %
The interactions that arise from an effective term in the Lagrangian allow that neutrino mass eigenstates can change due to the interaction, and this leads to a system of three propagation equations which we have solved in the cases of neutrino energies $E \lesssim 5\times 10^{9}$GeV, for which they can be decoupled.
We addressed the cases of point neutrino sources at different redshifts, and also the case of the a diffuse flux of neutrinos, considering the interactions with extragalactic and galactic DM. We have found that, either if the DM is self-conjugate or not, the interaction is sensible to the neutrino mass, affecting more to the most massive ones. This implies that the neutrino flavor composition is modified by the neutrino interactions, and, in particular, the cases of a normal or inverted hierarchy of the neutrino masses would be clearly different: dominated by electron neutrinos in the NH case, and by tau and muon neutrinos in the IH case. This is an interesting prediction to be considered as a possible mechanism for introducing departures from the normally expected $(f_e:f_\mu:f_\tau)\approx(1:1:1)$ neutrino flavor composition {(see Ref.\cite{palatable})}, as more and more data is accumulated by IceCube.
%

\section*{Acknowledgments}
We thank J.H. Peressutti for fruitful discussions about neutrino interactions.
We thank CONICET (PIP-2013-2015 GI 160), ANCyT (Pr\'estamo Bid PICT 2012-2621), and Univesidad Nacional de Mar del Plata (Proyecto 15/E755 EXA804/16) for financial support.

\end{document}